\documentclass[final,onefignum,onetabnum]{siamart190516}

\usepackage{amsfonts}
\usepackage{graphicx}
\usepackage{epstopdf}
\usepackage{algorithmic}

\usepackage{graphicx}
\usepackage{color}
\usepackage{amssymb}
\usepackage{amsfonts}
\usepackage{mathtools}
\usepackage{tikz} 
\usepackage{pgfplots}
\usepackage{bm,nicefrac}
\usepackage{listings}
\usepackage{extarrows}

\newtheorem{lem}{Lemma}[section]

\newtheorem{prop}{Proposition}[section]

\newtheorem{Th}{Theorem}[section]

\newcommand{\ovl}{\overline}

\newcommand{\ignore}[1]{}

\newcommand{\PP}{\mathbb{P}}

\newcommand{\Q}{\mathcal{Q}}

\newcommand{\blue}{\textcolor{black}}
\newcommand{\red}{\textcolor{black}}

\ifpdf
  \DeclareGraphicsExtensions{.eps,.pdf,.png,.jpg}
\else
  \DeclareGraphicsExtensions{.eps}
\fi

\newsiamremark{remark}{Remark}
\newsiamremark{hypothesis}{Hypothesis}
\crefname{hypothesis}{Hypothesis}{Hypotheses}
\newsiamthm{claim}{Claim}

\headers{Combinatorics of polymer models of early metabolism}{O. Weller-Davies, M, Steel, and J. Hein}

\title{Combinatorics of polymer-based models of early metabolism}

\author{Oliver Weller-Davies\thanks{Department of Statistics, Oxford University, Oxford UK 
  (\email{ocwellerdavies@gmail.com}, \email{hein@stats.ox.ac.uk}, \url{http://www.stats.ox.ac.uk/all-people/jotun-hein/}).}
\and Mike Steel\thanks{Biomathematics Research Centre, University of Canterbury, Christchurch, New Zealand 
  (\email{mike.steel@canterbury.ac.nz}).}
\and Jotun Hein\footnotemark[1]}

\usepackage{amsopn}

\ifpdf
\hypersetup{
  pdftitle={Combinatorics of polymer-based models of early metabolism},
  pdfauthor={O. Weller-Davies, M, Steel, and J. Hein}
}
\fi

\begin{document}

\maketitle

\begin{abstract}
  Polymer models are a widely used tool to study the prebiotic formation of metabolism at the origins of life. Counts of the number of reactions in these models are often crucial in probabilistic arguments concerning the emergence of autocatalytic networks. In the first part of this paper, we provide the first exact description of the number of reactions under widely applied model assumptions. Conclusions from earlier studies rely on either approximations or asymptotic counting, and we show that the exact counts lead to similar, though not always identical, asymptotic results. In the second part of the paper, we investigate a novel model assumption whereby polymers are invariant under spatial rotation. We outline the biochemical relevance of this condition and again give exact enumerative and asymptotic formulae for the number of reactions.

\end{abstract}



\section{Introduction}
The investigation of polymer models has lead to many foundational insights into the origins of life and metabolism. The protein-world hypothesis, for instance, was first motivated by the seminal work of Stuart Kauffman in the 1980s, who's arguments relied on a combinatorial analysis of the Binary Polymer Model, based on counts of the number of reactions and molecules present \cite{kau86}.
More recently, polymer models have appeared in dynamical molecular simulation flows \cite{hor12b}, theories of template-based metabolic emergence \cite{hor11} and in studies of the dynamics of autocatalytic cycles in partitioned chemical networks \cite{smi14}.

In this paper, we review existing polymer models (based on `oriented' polymers) and formalise some subtleties that have been hitherto overlooked. We note a distinction between two model assumptions that are often unknowingly conflated and investigate the consequences of this distinction. In both cases, we derive precise enumerative and asymptotic formulae for the number of reactions. We then explore another polymer model (`non-oriented') with combinatorial properties that do not appear to have been explored in previous work. We again derive exact and asymptotic formulae, and the consequences of these are also briefly discussed.

\section{Catalytic reaction systems and autocatalytic sets}
\label{sec:def}
In order to describe polymer models in a formal setting, we will first introduce the notion of a \textit{catalytic reaction system}. Formally, a catalytic reaction system (CRS) is a quadruple $\Q = (X, R, C, F)$ where:
\begin{itemize}
	\item $X$ denotes a set of \textit{molecule types}.
	\item $R$ denotes a set of {\em reactions}. Here a reaction $r \in R$ refers to the chemical process whereby a collection  $A$ of molecule types interact to produce  another collection $B$ of molecule types. Formally, $r$ can be regarded as simply the ordered pair $(A,B)$, though  we will generally use the more standard notation $A \rightarrow B$.  The molecule types in $A$ are referred to as the \textit{reactants} of the reaction and those in $B$ as the \textit{products} of the reaction.
	\item $C \subseteq X \times R$ denotes a \textit{catalysation} assignment, where if $(x, r) \in C$ we say that the molecule $x$ \textit{catalyses} the reaction $r$.
	\item $F \subseteq X$ denotes an ambient \textit{food set} of molecule types, which are assumed to be freely available in the environment.
\end{itemize} 

The CRS model allows us to formally describe the types of chemical landscapes in which \textit{autocatalytic} sets of reactions can materialise. Simply put, an autocatalytic set of reactions is one in which every reaction in the set is catalysed by the product of another reaction in the set. These autocatalytic reaction sets are said to play a key role in the origins of life \cite{hor18b}. They allow us to skirt the issues surrounding the \textit{error threshold paradox} (Eigen's paradox) which place limits on the lengths of genetic strings (proto-RNA) prior to the evolution of error-correcting enzymes \cite{smi95}. They also serve as a prime candidate for the study of the origins of metabolism; many metabolic networks (such as the \textit{reverse Krebs cycle}) can be neatly expressed as autocatalytic reaction sets \cite{hei10}.

The spontaneous emergence of such cycles, something argued to be necessary for the origins of life \cite{hor18b}, is an extensively studied topic. In the theoretical setting, this has perhaps been most prominently formalised by the notion of a \textit{Reflexively Auto-catalytic F-generated Set} (RAF set). The RAF set incorporates the autocatalytic condition with an additional constraint requiring the set to be derivable from its ambient chemical environment. Formally, with respect to a CRS $\Q = (X, R, C, F)$, we say a set $R' \subseteq R$ is a RAF set if and only if $R'$ is non-empty and the two following conditions hold \cite{ste13}:
\begin{itemize}
    \item[(RA)] \textit{Reflexively Auto-catalytic:} {Every reaction $r \in R'$ is catalysed by a molecule type $x$ that is either in the food set  $F$ or is the product of another reaction $r' \in R'$.}
    \item[(F)] \textit{F-generated}: {The reactions in $R'$ can be written in a linear order $r_0, r_1,\ldots,r_n$ such that for every reaction $r_i = (A_i, B_i) \in R'$, each reactant $x \in A_i$ is either in the food set or is the product of another reaction occurring earlier in the ordering; that is, $\forall x \in A_i$, $x \in F$ or $x \in B_j$ for some $j < i$.}
\end{itemize}

The study of RAF sets has now flourished into a field known as RAF Theory and has been applied in a number of different settings, particularly regarding the origin of metabolism. This includes both simulated and laboratory-based systems of early metabolism (\cite{ash04, vai12}, discussed in \cite{ste18}), as well as current \cite{sou15} and ancient \cite{xav19} metabolic systems. Polymer models are very frequently used within RAF Theory, and several mathematical results concerning the emergence of RAFs ultimately depend (often quite sensitively) on the number of reactions and molecule types, and the ratio of these two quantities. We briefly explore this further in section \ref{relesec}.

\begin{figure}[!t]
\begin{center}
\includegraphics[width=0.5\linewidth]{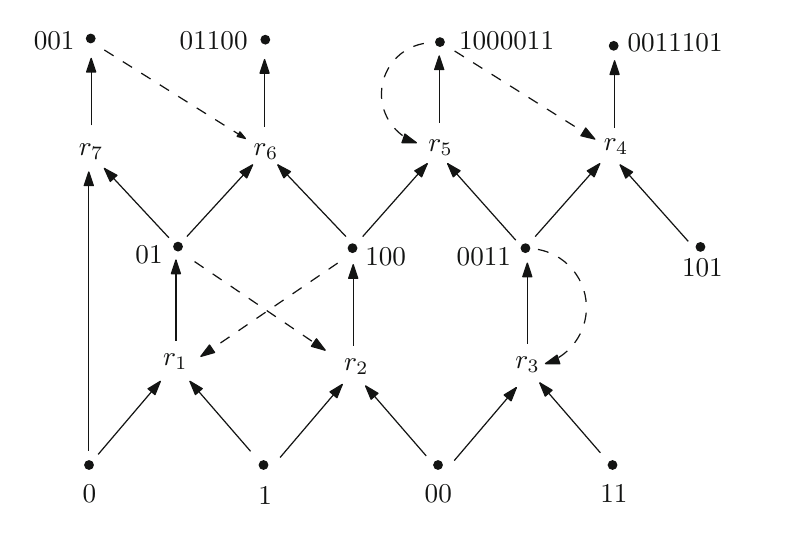} 
\caption{A simple CRS based on polymers over a two-letter alphabet (0,1), with a food set
$F = \{0, 1, 00, 11\}$ and seven reactions ($r_1$ -- $r_7$) (from  \cite{smi14}).  Dashed arrows indicate catalysis; solid arrows show reactants entering a reaction and products leaving. In this CRS, there are exactly four RAFs (defined below), namely $\{r_1, r_2\}, \{r_3\}, \{r_1, r_2, r_3\}$ and $\{r_1,r_2,r_3, r_5\}$.}
\label{fig1}
\end{center}
\vspace{-10pt}
\end{figure}

\section{Polymer Models}
A simple example of a polymer model is the Binary Polymer Model introduced by Stuart Kauffman in \cite{kau86}. This pioneering work was the perhaps the first to demonstrate the effectiveness of formal models in yielding insights into the origins of life; it largely motivated the protein-world hypothesis.

In the Binary Polymer Model, further developed more formally in \cite{mos05, ste00},  the set $X$ of molecule types consists of oriented polymers over the alphabet $\Sigma$, which contains $k$ symbols.  
More precisely, \blue{let $\Sigma^+$ be the (infinite) set of all words $w = x_1 \cdots x_n$ where $x_i \in \red{\Sigma}$  for each $i$ and $n\geq 1$ is variable (thus $w$ has length $|w|=n$).} For $n \geq 1$, let $X_n$ be the set of  words  $w \in \Sigma^+$ of length at most $n$ (i.e. $1 \leq |w|\leq n$).  Different biological contexts will dictate different configurations of $\Sigma$. For example, an alphabet $\Sigma = \{A,T,C,G\}$ (with $k = 4$) may aim to simulate RNA polymers. Alternatively, to represent catalytic networks prior to the arrival of RNA, an alphabet $\Sigma$ consisting of amino acids may be used, with the words representing polypeptides. In Kauffman's original 1986 paper, the alphabet $\Sigma = \{0, 1\}$ was investigated (hence the name Binary Polymer Model) \cite{kau86}.

In the situation described above,
the number of different molecule types $|X_n| = |\{ w \in \Sigma^+ : |w| \leq n \}|$ is just the sum $\sum_{i = 1}^{n} k^i$ (all non-empty words of length $\leq n$) and therefore: 
\begin{equation}
|X_n| = \frac{k^{n+1} - k}{k-1} \sim \frac{k^{n+1}}{k - 1}
\end{equation}
where here and below $\sim$ denotes asymptotic equivalence as $n$ grows (we regard $k$ as fixed throughout).
In other words,  $f(n) \sim g(n)$ if and only if  $\lim_{n \rightarrow \infty}{f(n) / g(n) = 1}$ (for details, see \cite{mos05}). 

In this model, reactions consist of two complementary types. The first is a  {\em ligation}  reaction where two polymers are concatenated:
$$w + w' \rightarrow ww'.$$
The second is a {\em cleavage} reaction where a polymer is split in two:
$$ww' \rightarrow w+w'.$$

Note that the cleavage and ligation reactions are reversals of each other and thus are in one-to-one correspondence.  We will therefore mostly concentrate on enumerating ligation reactions. We now highlight a subtle but important distinction in the following example. Consider the two ligation reactions:
\begin{equation}
\label{ww1}
ab + abab \rightarrow  ababab
\end{equation}
and 
\begin{equation}
\label{ww2}
abab +ab \rightarrow ababab.
\end{equation}
These two reactions have the same reactants and the same product; the only distinction is the order in which the reactants appear on the left. 
Thus it is tempting to regard (\ref{ww1}) and (\ref{ww2}) as the same reaction. We call this the \textsc{set} convention, and it has been tacitly assumed in several papers \cite{mos05, ste00}.  However, given that the polymers are oriented,
one might regard the first reaction above as attaching $ab$ to the left-hand end of the (oriented) polymer $abab$, whereas the second reaction is attaching $ab$ to the right-hand end of the
polymer $abab$. In this way, the reactions can regarded as different. We call this the \textsc{tuple} convention, and it has also been tacitly assumed in other papers involving the model, particularly with simulations. Note that the distinction between these two conventions vanishes in the non-oriented setting, which we will discuss in section \ref{nonorient}.

Let  $R^{\textsc{set}}_n$ and $R^{\textsc{tuple}}_n$ denote the set of ligation \red{and cleavage} reactions involving polymers of size less or equal to  $n$ under conventions \textsc{set} and \textsc{tuple}, respectively. The one-to-one correspondence between cleavage and ligation reactions holds for both conventions \red{(hence the factor of two in (\ref{twoass}))}. Calculating the size of $R^{\textsc{tuple}}_n$ is easy and was carried out in the earlier papers cited above.  
For $i\geq 2$, if we let $T(i)$ denote  the number of ordered triples of oriented polymers $(u,v,z)$ where $uv=z$ and $|z|=i$ for $u,v,z \in \Sigma^+$, 
then:
\begin{equation}
\label{eqs}
T(i) = (i-1)k^i,
\end{equation}
and we have:
\begin{equation} 
\label{twoass}
|R^{\textsc{tuple}}_n| = 2\sum_{i=2}^n T(i) = 2\left(\frac{k((n-1)k^{n+1} - nk^n + k)}{(k-1)^2}\right) \sim \frac{2nk^{n+1}}{k-1}
\end{equation}

This value for the number of reactions in polymer models is widely cited and underlies many results. Indeed, it is a core component of the main argument presented in the original 1986 Binary Polymer Model paper by Kauffman \cite{kau86} \blue{in which it is assumed that each molecule type has a fixed probability $p$ (independent of $n$) of catalysing each reaction. In that case,} the expected number of reactions that each molecule catalyses, $\mu_n$, grows exponentially in $n$, as \red{$\mu_n = p|R^{\textsc{tuple}}_n| = 2p(\frac{nk^{n+1}}{k-1})$}. Armed with this fact, Kauffman argued that autocatalytic sets were an emergent and inevitable consequence of particular types of polymer-\blue{based systems}.

However, in many papers such as \cite{mos05, ste00} and indeed Kauffman's original paper, the convention \textsc{set} (not \textsc{tuple}) is in fact assumed; the widely cited value above (\ref{twoass}) is therefore not perfectly accurate. Despite the widespread usage, no exact enumeration of $|R^{\textsc{set}}_n|$ has hitherto been given. 

To describe this, fix an alphabet $\Sigma$ of size $k$ and let $S(i)$ denote the number of unique unordered sets $\{u,v,z\}$ for $u,v,z \in {\Sigma}^+$ with $uv = z$ and $|z|= i$. (The function names $T(i)$ and $S(i)$ were chosen to convey the semantics of the \textsc{tuple} and \textsc{set} conventions, respectively.) We first state a lemma, which is a consequence of a more general result from Lyndon and Sch{\"{u}}tzenberger \cite{lyn62}. (Below, $w^i$ denotes $i$ repeated copies of the word $w$ joined together.)
\begin{lem}
\label{abba}
If $uv = vu$ for non-empty words $u$ and $v$, then $u$ and $v$ are powers of a common word; that is, there exists a word $w$ such that $w^i = u$ and $w^j = v$ for some $i,j \geq 1$. Therefore, $uv = w^{i+j}$. 
\end{lem}
\hfill$\Box$

This lemma allows us to characterise strings of the form $uv = vu
$ with $u, v \in \Sigma^+$. In particular, if $z$ is a word of the form $z = uv=vu$ for $u,v \in \Sigma^+$,  then $z$ is necessarily a word formed from a repeated sub-word. Words that can be decomposed in such a way are often referred to as \textit{periodic}, where their \textit{period} is taken to be the  smallest  sub-word that repeatedly joins to form the word. For instance, the word $abcabcabc$ has period $abc$. The period of a word, if it exists, must be unique. Conversely, words that cannot be formed from repeated substrings in such a way are often referred to as \textit{aperiodic} or \textit{primitive} words (e.g. $abcdef$ is aperiodic).

This leads to the following expression for $|R^{\textsc{set}}_n|$, where 
 $\mu$ denotes the (classical) M\"obius function for the partially ordered set of positive integers under division  \cite{sta11}, defined by:

\begin{equation}
\label{mobiii}
    \mu(x) = 
    \begin{cases*}
      1, & if $x = 1;$ \\
      0, & is $x$ is a square number; \\
      (-1)^r, & if $x$ is square-free with $r$ distinct prime factors. \\
    \end{cases*}
  \end{equation}

\begin{Th}
\label{main1}
Fix an alphabet $\Sigma$ of size $k$. We then have the following:

\begin{itemize}
\item[(i)] 
$|R^{\textsc{set}}_n|=2\sum_{i=2}^n S(i)$, where:
\begin{equation} \label{rMob}
S(i)=
    (i-1)k^i - \sum_{d|i, d<i}
    {
      {\left \lfloor \frac{i/d - 1}{2}\right \rfloor}
      \sum_{d'|d}{
          \mu\left(\frac{d}{d'}\right)k^{d'}
      }
    }
\end{equation}
  
  \item[(ii)]
$T(i) - S(i) = O(k^{i/3})$, where $T(i)$ is given in (\ref{eqs}),
and so $|R^{\textsc{set}}_n| \sim |R^{\textsc{tuple}}_n|$ as $n$ grows.
\end{itemize}
\end{Th}

To illustrate Theorem~\ref{main1}, consider the binary polymer model (i.e. $k=2$). By Part (i) of Theorem~\ref{main1} we have $S(2)=T(2)=4$ and $S(3) = 2.2^3 - 2 = 14$, while $T(3)= 16$. Thus  $|R_3^\textsc{set}| = 2(S(2)+S(3)) = 36$ and $|R_3^\textsc{tuple}| = 2(T(2)+T(3)) = 40$.  In this case, for the two ligation reaction pairs:
$$aa + a \rightarrow aaa, \mbox{  } a + aa\rightarrow aaa$$ 
$$bb + b \rightarrow bbb, \mbox{ } b+ bb \rightarrow bbb,$$ 
the reactions within each pair are counted separately by $T(3)$ but only once by $S(3)$ (which happens again, symmetrically, for the cleavage reactions).

\begin{figure}[ht]
\begin{center}
\includegraphics[width=0.8\linewidth]{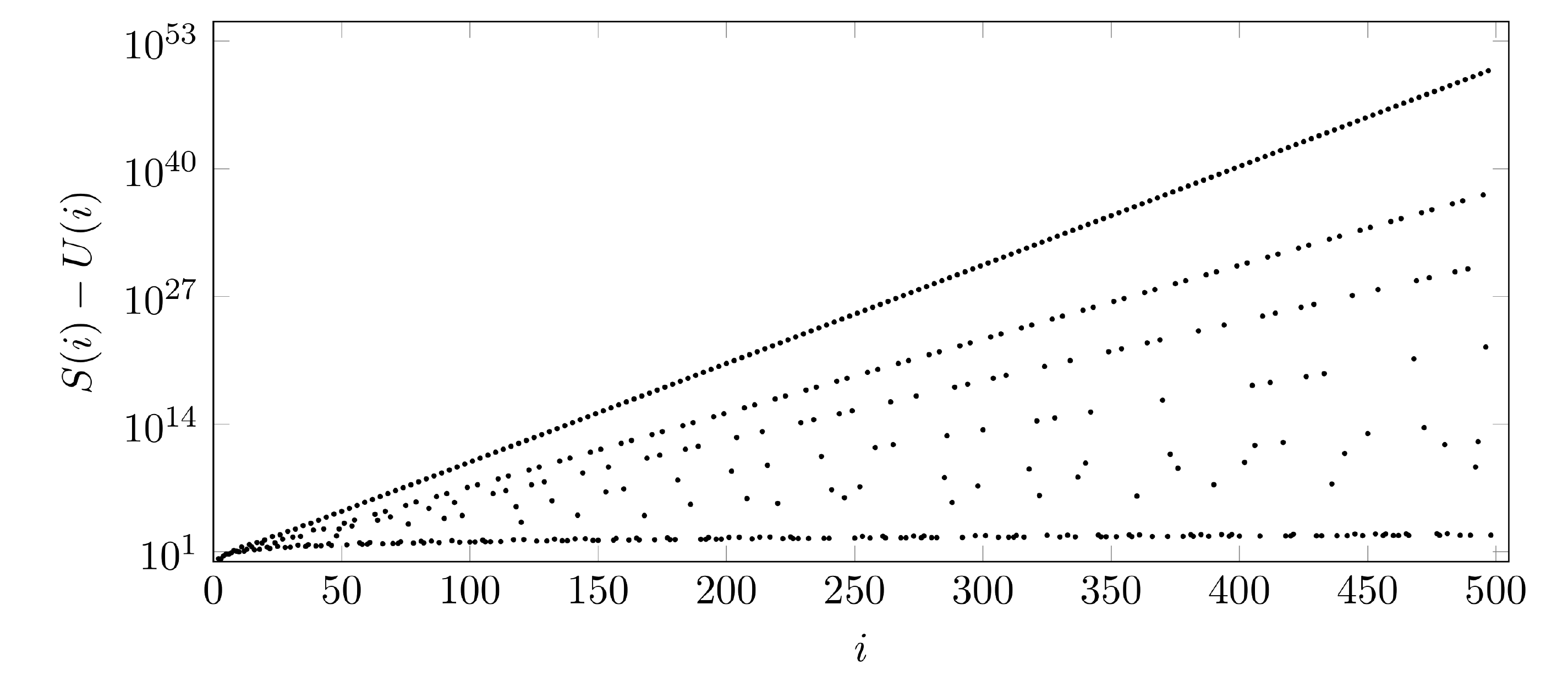} 
  \caption{The difference between $T(i)$ and $S(i)$ as $i$ increases from 1 to 500}
    \label{fig4}
\end{center}
\end{figure}

The  difference  $T(i) - S(i)$ quantifies the difference between the conventions when the product of a ligation reaction has size $i$. A graph of this difference is shown in Fig.~\ref{fig4}. 

{\em Outline of the proof of Theorem~\ref{main1}:}  We will prove Theorem~\ref{main1} in two parts, starting with Theorem~\ref{main1}(ii) and finishing with Theorem~\ref{main1}(i); the latter will be established through a series of four claims. We will first establish a Lemma relevant in both proofs.

 \begin{lem}
 Using convention \textsc{tuple} (not {\textsc{set}}) to count $S(i)$ results  in double-counts precisely whenever $z = uv = vu, u \neq v$ and $|z| = i$ for words $u,v,z \in \Sigma^+$.
 \label{uvvu}
 \end{lem}
{\em Proof.}
Observe that for $u,v,z \in \Sigma^+$ with $uv = z$ we double-count using convention \textsc{tuple} (instead of convention \textsc{set}) precisely when there are two distinct triples $(u,v,z)$ and 
$(u', v', z')$ with $uv = u'v' = z$ and the corresponding sets $\{u, v, z\} = \{u', v', z\}$ equivalent (hence $u + v \rightarrow z$ and $u' + v' \rightarrow z$ are counted twice in {\textsc{tuple}} and once together in {\textsc{set}}). Since $z$ is common to both sets, we have $\{u, v, z\} = \{u', v', z\}$ if and only if $\{u, v \} = \{u', v' \}$. However, since $(u, v, z)$ and $(u', v', z)$ are distinct, we also have $u \neq u'$ or $v \neq v'$. This gives us $\{u,v,z\} = \{u', v',z\}$ if and only if $u = v'$ and $v = u'$. Rearranging gives $uv = vu = z$. \\

\begin{figure}[ht]
\begin{center}
\includegraphics[width=0.4\linewidth]{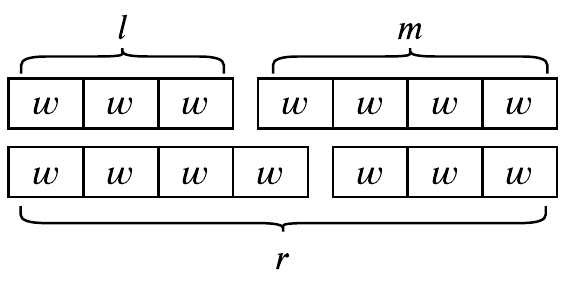} 
  \caption{Illustrated above is a word $W = wwwwwww$ of length $i$. We have $r = 7$ copies of $w$, each of size $|w| = i/r$. In the  \textsc{tuple} convention, the reactions $w^l + w^m \rightarrow w^{l+m}$ (top) and $w^m + w^l \rightarrow w^{l+m}$ (bottom) will be counted separately (twice). In the \textsc{set} convention, they will be counted together (once). For a given decomposition of $W$ (e.g. $W = wwwwwww$), dictated by a value $r = l + m$, we will have at most $k^{|w|} = k^{i/r}$ different choices for $w$ and $\frac{1}{2}(r-1)$ different values of $l, m$. The quantity $T(i) - S(i)$ is consequently bounded above by $\sum_{r=3}^i \frac{1}{2}(r-1) k^{i/r} = O(k^{i/3})$.}
  \label{double-count}
\end{center}
\end{figure}

{\em Proof of Theorem~\ref{main1} (ii):} Lemma \ref{uvvu} gives us that
$T(i) - S(i)$ represents the number of distinct sets of words $\{u, v\}$ for $u, v \in \Sigma^+$ with $uv = vu$ and $|uv| = i$ ($u + v \rightarrow uv$ and $v + u \rightarrow vu = uv$ are counted separately (twice) in $T(i)$ and only once together in $S(i)$). With this, Lemma~\ref{abba} further gives $T(i)-S(i)$ is the number of pairs of words of the form  $w^l, w^m$ with  $l, m \geq 1$ and $l < m$ where $(l+m)|w|=i$.
Let $r=l+m$, so $|w| = i/r$. For a given $r$, there will be $\frac{1}{2}(r-1)$ such values for $l, m$ with $l < m$ and $l + m = r$. Furthermore, since $l, m \geq 1$ and $l\neq m$, we have $r \geq 3$. With $|w|=i/r$ there are $k^{i/r}$ choices for values of $w$, and so $T(i)-S(i)$ is bounded above, as claimed, by:
  $$\sum_{r=3}^i \frac{1}{2}(r-1) k^{i/r} = O(k^{i/3})$$See Fig. \ref{double-count} for an illustration.
 
 The second claim in  Part (ii) now follows by applying Part (i) and Eqn.~(\ref{twoass}) to give: $$|R^{\textsc{set}}_n| = 2\sum_{i=2}^n S(i) \sim 2\sum_{i=2}^n T(i) = |R^{\textsc{tuple}}_n|$$ The asymptotic equivalence here holds because the sum of terms of order $k^{i/3}$ from $i=2$ to $i=n$ is of order $k^{(n+1)/3}$ and thus is asymptotically negligible compared with the rest of the sum. \hfill$\Box$

\bigskip 
{\em Proof of Theorem~\ref{main1} (i):} The proof consists of establishing a series of claims.

\bigskip

 \noindent {\bf Claim 1:} For each $i \geq 2$ there is a bijection between the set of periodic words of length $i$ and the set of aperiodic sub-words of length $d$ where $d | i$ and $d < i$. 

To see this, observe that every periodic word $w_i$ must have a unique period; the period is the \textit{smallest} sub-word that repeatedly joins to form the word. Further, that period must itself be aperiodic, and must be of length $d < i$ where $d | i$. Conversely, fix an aperiodic word $w_d$ of length $d$ with $d | i, d < i$. Then $w_d^{i / d} = w_i$ for some periodic word $w_i$ of length $i$ (and consequently $w_d$ is not the period of any other word of length $i$). This establishes Claim 1.

With this bijection, we can easily derive a recursive formula to count the number of aperiodic words of length $i$.

\bigskip

 \noindent {\bf Claim 2:} The number of aperiodic words $\hat{p}(i)$ of length $i \geq 1$ is given exactly by:
\begin{equation} \label{ap}
    \hat{p}(i) = k^i - \sum_{d|i, d < i} {\hat{p}(d)}
\end{equation}

To see this, observe that  the number of aperiodic words of length $i$ is clearly $k^i$ subtract the number of periodic words of length $i$. By Claim 1, we can count the number of periodic words of length $i \geq 2$ by counting the number of aperiodic words of length $d < i, d | i$. The number of aperiodic words of length $1$ is simply $k$. This gives Eqn. (\ref{ap}) and so establishes Claim 2. \\

With this, we now achieve a non-recursive formula. Rearranging (\ref{ap}) by adding the rightmost sum to both sides we get:
\begin{equation} \label{map}
    {k}^i = \sum_{d|i}{\hat{p}(d)}
\end{equation}
By applying the M\"obius inversion  formula we get:
\begin{equation}
        \hat{p}(i) = \sum_{d | i}{\mu\left(\frac{i}{d}\right)k^{d}}
\end{equation}
Recalling our original problem, where we double-count precisely whenever $uv = vu = z$, $u \neq v$ and $|z| = i$ for any $u,v,z \in \Sigma^+$, it follows from Lemma \ref{abba} that $z$ must be periodic. 

\bigskip

\noindent {\bf Claim 3:} When using convention \textsc{tuple} (rather than \textsc{set}) to count $S(i)$, the number of double-counts per periodic word of length $i$ with period length $d$ is given exactly by $\left \lfloor \frac{i/d -1}{2} \right \rfloor$.

To see this, let  $w_i$ be a periodic word of length $i$ and $w_d$ its period of length $d$. It follows that $w_{d}^jw_{d}^{i/d-j} = w_i$ for each $1 \leq j < i/d$. Using convention \textsc{tuple}, we will double-count precisely whenever $j < i/d - j$, as $j$ is symmetric for $i/d-j$. For an example, let $w_i=abcabcabcabc$. Here,  $w_i$ has a period $abc$ and there are $3$ ways to partition $w_i$ with units $abc$ using convention \textsc{tuple}:\begin{center}
    (1) $\blue(abc,   {abc}{abc}{abc},   w_i\blue)$
    \item (2) $\blue(abcabc,  {abc}{abc},  w_i\blue)$
    \item (3) $\blue(abcabc{abc},  {abc},  w_i\blue)$
\end{center}

However,  (1) and (3) are the same \blue{\red{when considered} as} sets and therefore double-counted in the unordered version. Combining $j < i/d - j$ with $j \geq 1$ gives us $\left\lfloor\frac{i/d-1}{2} \right\rfloor$ double-counts. This establishes Claim 3. \\

\bigskip

\noindent{\bf Claim 4:}  $S(i)$ is given by:
 $$S(i) = (i-1)k^i - \sum_{d|i, d<i}
    {
      {\left \lfloor \frac{i/d - 1}{2}\right \rfloor}
      \sum_{d'|d}{
          \mu\left(\frac{d}{d'}\right)k^{d'}
      }
    }$$

To establish  Claim 4 we use convention \textsc{tuple} (which leads to Eqn.~(\ref{eqs})) to give  $(i-1){k}^i$ and then subtract the double-counts
to give our desired convention \textsc{set}.  From Claim 3, each periodic word of length $i$ and period $d$ will incur exactly ${\left \lfloor \frac{i/d - 1}{2}\right \rfloor}$ double-counts. Further, from Lemma \ref{uvvu}, we double-count exclusively when this is the case. This gives us the formula:
$${   S(i) = 
    (i-1){k}^i - \sum_{d|i, d<i}
    {
      {\left \lfloor \frac{i/d - 1}{2}\right\rfloor}
      \hat{p}(d)
  }
}$$
Using the explicit formula for $\hat{p}(d)$, we arrive at Eqn. (\ref{rMob}).
This establishes Claim 4 and thereby Part (i). 

\hfill$\Box$

\bigskip

\subsection{Non-oriented polymers}
\label{nonorient}
The choice to consider polymers as oriented is not only computationally convenient, but it also has some connection with bio-molecular sequence data, where (for example) DNA polymerase has an orientation (from the 5$'$ to 3$'$ end of the sequence). 
However, when the symbols (monomers) that form the alphabet $\Sigma$ have no pre-stated orientation, then a polymer such as $accb$ can be considered identical to $bcca$, since one can rotate the first molecule in space to obtain the second. Thus, in such a setting, each oriented polymer $w$ can be viewed as equivalent to its `reverse' oriented polymer $w^{-}$, obtained by reversing the order of the symbols in $w$.
Note that a polymer is equal to its associated reverse polymer  precisely if it is a palindromic polymer (e.g. $acca$).  In this way, the set $\Sigma^+$ is 
partitioned into pairs and singleton classes under the equivalence relation $w \sim w'$ if $w' = w$ or $w'=w^{-}$. We refer to these equivalence classes as {\em non-oriented polymers} and will write $\overline{w}$ for the equivalence class of $w$. 
Let $\overline{X}_n = \{\overline{w} : w \in \Sigma^+ \text, |w| \leq n\}$ denote the set of non-oriented polymers (equivalence classes) for words in $\Sigma^+$.

\bigskip

\noindent\fbox{%
    \parbox{\textwidth}{%
\noindent {\em Example:} 

For $\Sigma= \{a,b\}$, we have:  $|\overline{X}_2| = 5$, since $$\overline{X}_2 = \{\{a\}, \{b\}, \{aa\}, \{bb\}, \{ab, ba\}\}.$$ 
Similarly, $|\overline{X}_3| = 11$, since: $$\overline{X}_3 = \{\{a\}, \{b\}, \{aa\}, \{bb\}, \{ab, ba\}, \{aaa\}, \{bbb\}, \{aba\}, \{bab\}, \{abb, bba\}, \{aab, baa\}\}.$$
By contrast, $|X_2|=6$ and $|X_3| = 14$.
}}

\begin{lem}
\label{lemolem}
$$
|\overline{X}_n| =
\frac{k}{2(k-1)} \left(
k^n +
k^{\lceil n/2 \rceil} + k^{\lfloor n/2 \rfloor} - 3
\right)
$$



In particular, $|\overline{X}_n| \sim \frac{1}{2} |X_n|$ as $n$ grows.
\end{lem}
{\em Proof:} Let $p_i$ (respectively, $n_i$) denote the number of palindromic (respectively non-palindromic) oriented polymers of length $i$.
We have:
$$|\overline{X}_n|  = \sum_{i=1}^n \frac{1}{2} n_i  + \sum_{i=1}^n p_i = \frac{1}{2} \left(\sum_{i=1}^{n} k^i + \sum_{i=1}^n p_i\right)$$
where the second equality follows from the identity:
 $n_i = k^i - p_i$. 
Now,  $p_i$ is equal to  $k^{i/2}$ when  $i$ is even and is equal to $k^{(i-1)/2} \times k = k^{(i+1)/2}$ when $i$ is odd:

$$ |\overline{X}_n| = \frac{1}{2} \left(\sum_{i=1}^{n} k^i + \sum_{i=1}^n p_i\right) = \frac{1}{2} \left(\sum_{i=1}^{n} k^i + \sum_{i=1}^{\lfloor n/2 \rfloor} p_{2i} + \sum_{i=1}^{\lceil n/2 \rceil} p_{2i-1}  \right)$$

$$= \frac{1}{2} \left(\sum_{i=1}^{n} k^i + \sum_{i=1}^{\lfloor n/2 \rfloor} k^i + \sum_{i=1}^{\lceil n/2 \rceil} k^i  \right)$$

Applying the geometric sequence identity
$\sum_{j=1}^m x^j = x(x^{m} -1)/(x-1)$ establishes the Lemma.
\hfill$\Box$

\bigskip

Next, we consider the enumeration of reactions  involving non-oriented polymers. 
Consider the following  ligation reaction: 
\begin{equation}
\label{rr}
r: \overline{u} + \overline{v} \rightarrow \overline{z},
\end{equation} 
where $u, v$ and $z$ are oriented polymers.
Note that for any two oriented polymers $x,y$ we always have $(xy)^- = y^- x^-$, and the reaction $r$ in (\ref{rr}) holds whenever  $z$ is one of the following:
$$z = uv \mbox{ } (= (v^-u^-)^-)$$ 
$$z = uv^-\mbox{ }  (= (vu^-)^-)$$ 
$$z = u^-v \mbox{ }  (=(v^-u)^-)$$ 
$$z = u^-v^- \mbox{ }  (=(vu)^-)$$
Thus $\overline{u}$ and $\overline{v}$ could ligate to form up to four different non-oriented polymers, depending on how many of these four possible $z$ are equivalent to each other (allowing reversals). 
There are four cases to consider:
\begin{itemize}
\item[(i)]
$u$ and $v$ are distinct (i.e. $u \neq v, u \neq v^{-}$) non-palindromic polymers . In this case, there are four distinct choices of $z$ (no two of which are a reversal of each other),  so we obtain four distinct reactions of the type in (\ref{rr}).
\item[(ii)]If  $u,v$ are both non-palindromic and $u=v$ or $u=v^-$  we obtain three distinct choices of $z$ and so two additional distinct reactions of the type in (\ref{rr}). 
\item[(iii)] If one of $u$ or $v$ is a palindromic polymer and the other is not, 
we obtain two distinct choices of $z$ and so two distinct reactions of the type in  (\ref{rr}).
\item[(iv)] If both $u$ and $v$ are palindromic polymers, then all choices of $z$ are equivalent and so there is just one reaction of the type in (\ref{rr}).
\end{itemize}

\noindent\fbox{%
    \parbox{\textwidth}{%
\noindent {\em Example:}   

As an example of Case (i), consider the reaction of type (i):
$\overline{ab} + \overline{baa} \rightarrow \overline{abbaa}.$
The reactants on the left, give rise to three additional distinct reactions, namely:
$$\overline{ab} + \overline{baa} \rightarrow \overline{abaab}, \mbox{ } 
\overline{ab} + \overline{baa} \rightarrow \overline{babaa}, \mbox{ and } 
\overline{ab} + \overline{baa} \rightarrow \overline{baaab}.$$

Next, consider  the reaction of Type (ii):
$\overline{ab} + \overline{ba} \rightarrow \overline{abba}.$
The reactants on the left also give rise to: 
$$\overline{ab} + \overline{ba} \rightarrow \overline{baba} \mbox{ and }  \overline{ab} + \overline{ba} \rightarrow \overline{baab}.$$

Next, consider the reaction of Type (iii):
$\overline{aa} + \overline{ab} \rightarrow \overline{aaab}.$
The reactants on the left give rise to one additional reaction, namely 
$\overline{aa} + \overline{ab} \rightarrow \overline{aaba}.$

Finally,  the reaction of Type (iv)
$\overline{aa} + \overline{bab} \rightarrow \overline{aabab}$
gives rise to no further reactions. 

}
}
\bigskip

By applying these cases, once can,  in principle, count the number $N_{r,s}$ of ligation reactions in which two non-oriented polymers of size $r$ and $s$ are combined to give a non-oriented polymer of length $i=r+s$. Let ${\ovl{p}}_m, \ovl{n}_m$ denote the number of non-oriented polymers of length $m$ that are palindromic and non-palindromic, respectively. 
First, suppose that  $r<s$, so that Case (ii) cannot arise. We then have:
$$N_{r,s}= (2{\ovl{n}}_r + {\ovl{p}}_r)\cdot (2{\ovl{n}}_s+ {\ovl{p}}_s)$$
The case where $r=s$ (in which case, $i= 2r$ is even) requires a separate description as Case (ii) can now also arise. For the case $r=s$, we have the following identity (grouped by the contribution from each case): $$N_{r,r}=4\binom{n_r}{2} + 3{\ovl{n}}_r+ 2{\ovl{n}}_r{\ovl{p}}_r + \left(\binom{\ovl{p}_r}{2} + \ovl{p}_r \right)$$ This simplifies to:$$
N_{r,r} = 2{\ovl{n}}_r({\ovl{n}}_r-1) + 3{\ovl{n}}_r+ 2{\ovl{n}}_r{\ovl{p}}_r + {\ovl{p}}_r({\ovl{p}}_r+1)/2$$

\noindent\fbox{%
    \parbox{\textwidth}{
\noindent {\em Example:}   

Consider the case  $r=s=2$.  We have ${\ovl{n}}_2=1, {\ovl{p}}_2=2$ and so $N_{2,2}=10$.
These correspond to the following reactions, classified according to the cases described above. For Case (i) we have
 $2{\ovl{n}}_2({\ovl{n}}_2-1)=0$, (i.e. no reactions are possible in this case).

For Case (ii),  $3{\ovl{n}}_r=3$:  $\overline{ab}+\overline{ab} \rightarrow \overline{abba},\mbox{ } \overline{ab}+\overline{ab} \rightarrow \overline{abab},\mbox{  } \overline{ab}+\overline{ab} \rightarrow \overline{baab}.$

For Case (iii), $2{\ovl{n}}_2{\ovl{p}}_2 = 4$:
$$\overline{aa}+\overline{ab} \rightarrow \overline{aaab},  \mbox{ } \overline{aa}+\overline{ab} \rightarrow\overline{aaba},$$
$$\overline{bb}+\overline{ab} \rightarrow \overline{abbb},  \mbox{ } \overline{bb}+\overline{ab} \rightarrow \overline{babb}$$

For Case (iv),  ${\ovl{p}}_2({\ovl{p}}_2+1)/2 = 3$:
$\overline{aa}+\overline{aa} \rightarrow \overline{aaaa}, \mbox{  }  \overline{bb}+\overline{bb} \rightarrow \overline{bbbb}, \mbox{  } \overline{aa}+\overline{bb} \rightarrow \overline{aabb}.$
    }
}

\bigskip

The final case is when $r > s$. However, this case is symmetric to $r < s$:

\noindent\fbox{%
    \parbox{\textwidth}{
\noindent {\em Example:}   

Consider in $N_{2,3}$ the count of the reaction: $$\ovl{ab} + \ovl{aab} \rightarrow \ovl{abaab}$$And consider in $N_{3,2}$ the count of the reaction: $$\ovl{aab} + \ovl{ab} \rightarrow \ovl{abaab}$$Since the polymers are non-oriented, we can rotate $\ovl{aab}$, $\ovl{ab}$ and $\ovl{abaab}$ (from $N_{3,2}$) to achieve the reaction $\ovl{baa} + \ovl{ba} \rightarrow \ovl{baaba}$. However, as the polymers have no pre-stated orientation, the reaction $\ovl{baa} + \ovl{ba} \rightarrow \ovl{baaba}$ exactly the same as the reaction $\ovl{ab} + \ovl{aab} \rightarrow \ovl{abaab}$ (from $N_{2,3}$), only rotated. In a physical interpretation that is invariant under spatial rotation, the two reactions are identical. The counts of reactions in $N_{r,s}$ and $N_{s,r}$ are therefore in one-to-one correspondence. 
    }
}
\bigskip{}

The total number of ligation reactions in the non-oriented setting in which the product polymer has size exactly $i$ is given by $\sum_{r= 1}^{\lfloor i/2\rfloor}N_{r,i-r}$. We index up to ${\lfloor i/2\rfloor}$ to avoid counting identical reactions when $r > i - r$ (as illustrated in the example above). In principle, this provides an explicit (albeit complicated) expression for the total number of reactions for all polymers with length at most $n$.
Here, we just report the asymptotic behaviour as $n$ grows. 
We have the following result: if we compare the non-oriented case to the oriented, the number of reactions  is (asymptotically) half, and the ratio of reactions to molecule types remains the same.

\begin{prop}
\mbox{}
\label{prol}
Let $\overline{R}_n$ denote the set of \red{ligation and cleavage} reactions involving non-oriented polymers that have length at most $n$.
As $n$ grows and with $*=\textsc{tuple}, \textsc{ set}$, we have:
$|\overline{R}_n| \sim \frac{1}{2}|R^*_n|$
and $|\overline{R}_n| /|\overline{X}_n| \sim  |R^*_n|/|X_n|$. 
\end{prop}
{\em Proof:} We first note that the total number of reactions $|\overline{R}_n|$ in the non-oriented setting is given by:
$$|\overline{R}_n| = 2\sum_{i= 2}^n \sum_{r= 1}^{\lfloor i/2\rfloor}N_{r,i-r}$$
This follows as ligation and cleavage reactions are in one-to-one correspondence \red{(hence the factor of two)} and \red{$\sum_{r= 1}^{\lfloor i/2\rfloor}N_{r, i - r}$} is the number of ligation reactions in the non-oriented setting in which the product polymer has size exactly $i$.

We now note that $\ovl{p}_i = p_i$ and $\ovl{n}_i = \frac{1}{2}n_i$ for all $i$. In other words, the number of non-oriented palindromic polymers of length $i$ equals the number of oriented ones, and the number of non-oriented non-palindromic polymers of length $i$ is half the number of oriented ones. Since $p_i = O(k^{i/2})$, then $\ovl{p}_i = O(k^{i/2})$, and as $n_i = k^i - p_i$, then $n_i \sim k^i$ and $\ovl{n}_i \sim \frac{1}{2}k^i$. With this, it follows that $N_{r, i-r} \sim 4\ovl{n}_r\ovl{n}_{i - r}$ when $r < \lfloor i/2\rfloor$ and $N_{r, i-r} = 2\ovl{n}_r\ovl{n}_{i - r}$ when $r = \lfloor i/2\rfloor$. Substituting in $\ovl{n}_i \sim \frac{1}{2}k^i$, we conclude that $\ovl{n}_r\ovl{n}_{i - r} \blue{\sim} \frac{1}{4}k^i$, and therefore:

$$|\overline{R}_n| = 2\sum_{i= 2}^n \sum_{r= 1}^{\lfloor i/2\rfloor}N_{r,i-r}\sim 2\sum_{i= 2}^n \sum_{r=1}^{\lfloor i/2\rfloor} k^i \sim \sum_{i= 2}^n{i}k^i \sim \sum_{i= 2}^nT(i) = \frac{1}{2}|R^{\textsc{tuple}}_n| \sim \frac{1}{2}|R^{\textsc{set}}_n|$$

The second result in the proposition holds by $|\overline{R}_n| \sim \frac{1}{2}|R^*_n|$ and Lemma \ref{lemolem}.
\subsection{Consequences for the emergence of RAFs in polymer systems}
\label{relesec}

We now turn to the relevance of the enumeration of the various (related) classes of polymers and ligation-cleavage reactions to the degree of catalysis required for the emergence of RAFs.
Recall that a CRS $\Q=(X, R, C, F)$ consists not only of a molecule set $X$ and a set of reactions $R$,  but also an assignment of catalysis $C$ and a food set $F$. 
To emphasise that $\Q$ depends on the maximum polymer length $n$, we will often write this as $\Q_n$.  
In polymer models, $F$ is typically taken to be all words of length at most  $t$ (where $t$ is usually small and independent of $n$ (e.g. $t=2$)). 
As for catalysis, this is assigned randomly, and various models have been proposed. The simplest (dating back to Kauffman ~\cite{kau86, kau93}) assumes that each molecule
catalyses each reaction with a constant probability $p=p_n$ (sometimes dependent on $n$) and  that such events are independent across all pairs $(x,r) \in X \times R$.

Thus each molecule type catalyzes an expected number $\mu_n = p_n |R|$ of reactions.  Notice that if $p_n$ is independent of $n$, then $\mu_n$  grows exponentially with $n$. However, \red{across a range of proposed models of catalysis},
it turns out that $\mu_n$ needs only to grow linearly with $n$ for RAFs to arise with high probability \cite{mos05}. Moreover, there is a sharp transition here in the following sense (following from results 
in \cite{mos05}). In a polymer model with oriented polymers and under either convention \textsc{set} or \textsc{tuple} above, we have for any $\epsilon>0$:

$$\mu_n = n^{1-\epsilon} \Longrightarrow \lim_{n \rightarrow \infty} \PP(\mbox {there exists a }  RAF \mbox{ for }  \Q_n) = 0;$$
$$\mu_n = n^{1+\epsilon} \Longrightarrow \lim_{n \rightarrow \infty} \PP(\mbox {there exists a }  RAF \mbox{ for } \Q_n) = 1.$$
This linear transition has been observed in  numerous simulation studies (first in \cite{hor04}). 
A more fine-grained analysis (also from \cite{mos05}) shows that if we write $\mu_n=\lambda n$ then, for any fixed $n$:
\begin{equation}
\label{ppeqs}
\PP(\mbox {there exists a }  RAF \mbox{ for }  \Q_n) \rightarrow  \begin{cases}
0, &  \mbox{ as $\lambda \rightarrow 0$},\\
1, & \mbox{ as $\lambda$ grows.}
\end{cases}
\end{equation}

The linear dependence of catalysis rate on $n$ in the transition from having no RAF to having a RAF in the oriented polymer setting is essentially because  $n$ is asymptotic to the ratio of the number of reactions divided by the number of molecule types (i.e. $|R_n|/|X_n|$).  For more general `polymer-like' systems (including non-oriented polymers), there is an analogue of Eqn. (\ref{ppeqs}) in Theorem 1 of \cite{smi14}, where again the ratio of reactions to molecule types plays a key role.   This is the main reason why it is important to have an asymptotic  measure of the size of these sets.

\blue{The main significance of our results is that we have provided exact expressions for the ratio of reactions-to-polymers for the three models considered (TUPLE, SET and non-oriented). These in turn lead to the following asymptotics for these three models:
\begin{equation}
\label{asym}
\frac{|R_n^\textsc{tuple}|}{|X_n|} \sim 2n; \mbox{ }  \frac{|R_n^\textsc{set}|}{|X_n|} \sim \red{2}n; \mbox{ }  \frac{|\overline{R}_n|}{|\overline{X}_n|} \sim n,
\end{equation}
from which it follows that Eqn.~(\ref{ppeqs}) holds for all three models. The \red{third} asymptotic result in Eqn.~(\ref{asym}) is new, while the \red{first} two relied on arguments that overlooked the distinction between tuple and set.
The exact expressions for these three ratios given in our paper will allow for more precise estimates (for a given $n$) of the catalysis rates that lead to RAFs in these three models.} \red{Moreover, although the TUPLE and SET conventions are asymptotically identical, they do exhibit significant differences for small values of $n$ (e.g. $|R_3^\textsc{set}| = 36$ and $|R_3^\textsc{tuple}| = 40$). It has been shown that numerical simulations of polymer models often depend extremely sensitively on the value for the number of reactions (see \cite{hor04}, though equation (\ref{ppeqs}) hints at this). Furthermore, since the number of reactions and molecules both increase exponentially with $n$ (as discussed), numerical simulations are prevented from exploring large values of $n$, making the differences between the conventions especially relevant. For these reasons it is helpful to understand and highlight the distinction between the models and have exact formulas for the number of reactions in each case.}




\section{Concluding comments}
\label{sec:conc}

In this paper we provided the first exact formula for the number of reactions in polymer models and established asymptotic results that verify earlier findings based on heuristic approximations. We also explored a new model assumption which has biochemical relevance and described the implications of our results for the catalysis levels required for RAFs to form.
In future work, it would be interesting to enumerate  molecule types and reactions involving more complex CRS systems that are not based soley on polymers. A particularly relevant setting would involve enumerating the set of molecule types that involve molecules consisting of up to $n$ atoms chosen (with repetition)  from the six elements essential to life (carbon, hydrogen, oxygen, nitrogen, phosphorous and sulpher) and connected by covalent bonds into a connected graph (i.e. molecule) so as to respect valency.


\bibliographystyle{siamplain}
\bibliography{RAF}
\end{document}